\documentclass[journal]{IEEEtran}
\usepackage{epsfig,latexsym,amssymb,amsmath,graphics}
\usepackage{graphicx}
\usepackage{amssymb,dsfont,amsthm}
\usepackage{cite,color}
\usepackage{url}
\usepackage{booktabs}
\usepackage{algorithm, algorithmicx}
\usepackage{algpseudocode}
\usepackage{stfloats}
\usepackage{verbatim}
\usepackage{epstopdf}
\usepackage{subfig}

\definecolor{ColorRed}{rgb}{0,0,0}

\allowdisplaybreaks[4]

\newtheorem{theorem}{Theorem}

\begin{document}
\title{Map-Assisted Constellation Design for mmWave WDM with OAM in Short-Range LOS Environment}

\author{Yuan Wang, Chen Gong, Nuo Huang, and Zhengyuan Xu
	\thanks{This work was supported by the National Natural Science Foundation of China (Grant No. 62171428) and (Grant No. 62101526), Key
		Program of National Natural Science Foundation of China (Grant No. 61631018), Key Research Program of Frontier Sciences of CAS (Grant No. QYZDY-SSW-JSC003), and the Fundamental Research Funds for the Central Universities (Grant No. KY2100000118).
		
		Yuan Wang, Chen Gong, Nuo huang, and Zhengyuan Xu are with Key Laboratory of Wireless-Optical Communications, Chinese Academy of Sciences, University of Science and Technology of China, Hefei, Anhui 230027, China.
		Email: wangy001@mail.ustc.edu.cn; \{cgong821,huangnuo,xuzy\}@ustc.edu.cn.
		
		}}

\maketitle

\begin{abstract}

We consider a system that integrates positioning and single-user millimeter wave (mmWave) communication, where the communication part adopts wavelength division multiplexing (WDM) and orbital angular momentum (OAM). This paper addresses the multi-dimensional constellation design in short-range line-of-sight (LOS) environment, with stable communication links. We propose a map-assisted method to quantify the system parameters based on positions and reduce real-time computing overhead. We explore the possibility of using a few patterns in the maps, and investigate its performance loss. We first investigate the features of OAM beams, and find that the link gain ratio between any two sub-channels remains unchanged at some postions. Then, we prove that a fixed constellation can be adopted for the positions where the link gain matrices are sufficiently close to be proportional. Moreover, we prove that the system can adopt a fixed power vector to generate a multi-dimensional constellation if the difference between fixed power vector and optimal power vector is small. Finally, we figure out that the constellation design for all receiver locations can be represented by a few constellation sets.


\end{abstract}


\section{Introductions}

Currently, intelligent terminals are becoming more and more widely deployed, which provide not only positioning, tracking and other sensing services, but also high-speed communication services \cite{qi2020integrated}. 
However, with the development of communication technologies, the demand for spectrum resource is increasing. Such trend calls for high-frequency spectrum like millimeter wave (mmWave) and Tera Hertz (THz) \cite{ 6005345, sun2016investigation }. In this work, we mainly consider mmWave spectrum. Compared with sub-6GHz communication, mmWave communication can occupy wider spectrum resource (30-300 GHz) to guarantee high-speed transmission \cite{8344116}. On the other hand, orbital angular momentum (OAM), as a new degree of freedom, is applied to mmWave systems to further increase the transmission rate \cite{7968418, mmWaveOAM2014, ren2017line, Travelingwave2015}.

Due to the characteristics of beam diffusion and coaxial transmission, OAM is only suitable for short-range line-of-sight (LOS) communication \cite{ren2017line, yan201632, sasaki2018experiment, zhang2016orbital}. Moreover, higher spectrum will lead to larger propagation, penetration and reflection losses \cite{ liu2016line,8046024}, which further limits the application to short-range LOS environment.
In most existing mmWave systems, data processing is performed only in a single \textcolor{ColorRed}{continuous spectrum, even with wide bandwidth \cite{7445130, 7959626}, and the advantage of mmWave with abundant continuous or separate spectrum can be further utilized \cite{8241348, 8252751}. Benefiting} from the large link losses, mmWave systems can aggregate the licensed or unlicensed broad spectrum with small external interference to increase transmission rate.
In general, broad spectrum communication is combined with wavelength division multiplexing (WDM) technology, which adopts multiple carrier wavelengths to transmit data streams \cite{8765239}. 



For mmWave spectrum communication in sparsely scattered environment, such as conference room and workshop, the influence of multi-path is limited and the LOS path is dominant \cite{wang2018survey}. Meanwhile, the mmWave propagation shows an apparent quasi-optical property according to experimental measurement \cite{maltsev2009experimental}. Hence, the mmWave short-range LOS link can be considered as a pure LOS channel \cite{liu2016line}. Based on the above feature, the link conditions of mmWave communication in short-range LOS environment are only related to the positions of transmitter and receiver \cite{chen2020multi}. Generally, high accuracy positioning is necessary for some short-range intelligent systems, such as industrial internet and indoor robot. If OAM is adopted, this position information can be used to align the receiving antenna axis with transmitting antenna axis. Certainly, the systems can directly utilize OAM signals to locate and communicate \cite{chen2020multi, chen2018beam}.

Motivated by the quasi-static feature and positioning-communication integration, we attempt to configure the communication system based on the transceiver's position. Due to the stable gains for short-range LOS links, the optimal system parameters are determined when the position information is obtained. Hence, we construct a look-up table, which is termed as map method, to store the system parameters of all interested positions \cite{8865431}. Adopting map-assisted approach has two potential advantages: (1) For the system whose positioning service is active, the transmitter and receiver always have position information. If a communication request appears, the system can directly configure the transmitter and receiver by adopting the parameters stored in the map. Since there is no need for the time and transmission symbol overhead of channel estimation, channel feedback and real-time optimization, the delay of initial communication establishment will be reduced. (2) The system can solve computationally intensive optimization problems in an offline manner and store the optimization results into the map, which significantly relaxes the real-time computational pressure.

This work considers the mmWave WDM with different OAM modes under short-range LOS scenario. We mainly focus on the multi-dimensional constellation design problem which primarily improves the detection performance. Considering map-assisted communication systems, we analyze the metric loss when the system lies at a certain position but adopts the configuration parameters of another position. We consider multi-dimensional constellation design using minimum Euclidean distance (MED) as the performance criterion. We investigate the characteristics of OAM beams with different wavelengths and modes, and find that the link gain matrices can be proportional at certain positions. To characterize the feasibility of assigning few constellation sets to the entire interested positions, we prove that for any position 2, the normalized MED difference of using the constellation of position 1 can be sufficiently small if the link gain matrices of positions 1 and 2 are sufficiently close to be proportional. Thus, a fixed constellation can be adopted for these two positions. We also consider the system with fixed power allocation, and prove that the performance loss at certain positions is small when using one fixed power allocation scheme to generate a multi-dimensional constellation. The claim of only a few constellation patterns is supported by the numerical results.

The remainder of this paper is organized as follows.  In Section \ref{sec.system_model}, we provide the system model, elaborate the features of OAM beams and display the configuration process of map-assisted systems. In Section \ref{section.Cons_design}, we investigate the multi-dimensional constellation design, analyze the features of link gain ratio, figure out that the same link gain ratio leads to the same constellation, and that the constellation design can be represented by only a few constellation sets in the map. Moreover, we also investigate the effect of fixed power allocation on constellation design. Section \ref{section.NumericalResults} presents some numerical results to verify our theoretical analysis. Finally, Section \ref{section.conclusion} concludes this paper.

\graphicspath{{figures/Fig_SysModel/}}

\section{System model} \label{sec.system_model}

\begin{figure*}[htbp]
	\centering
	\includegraphics[width=5in]{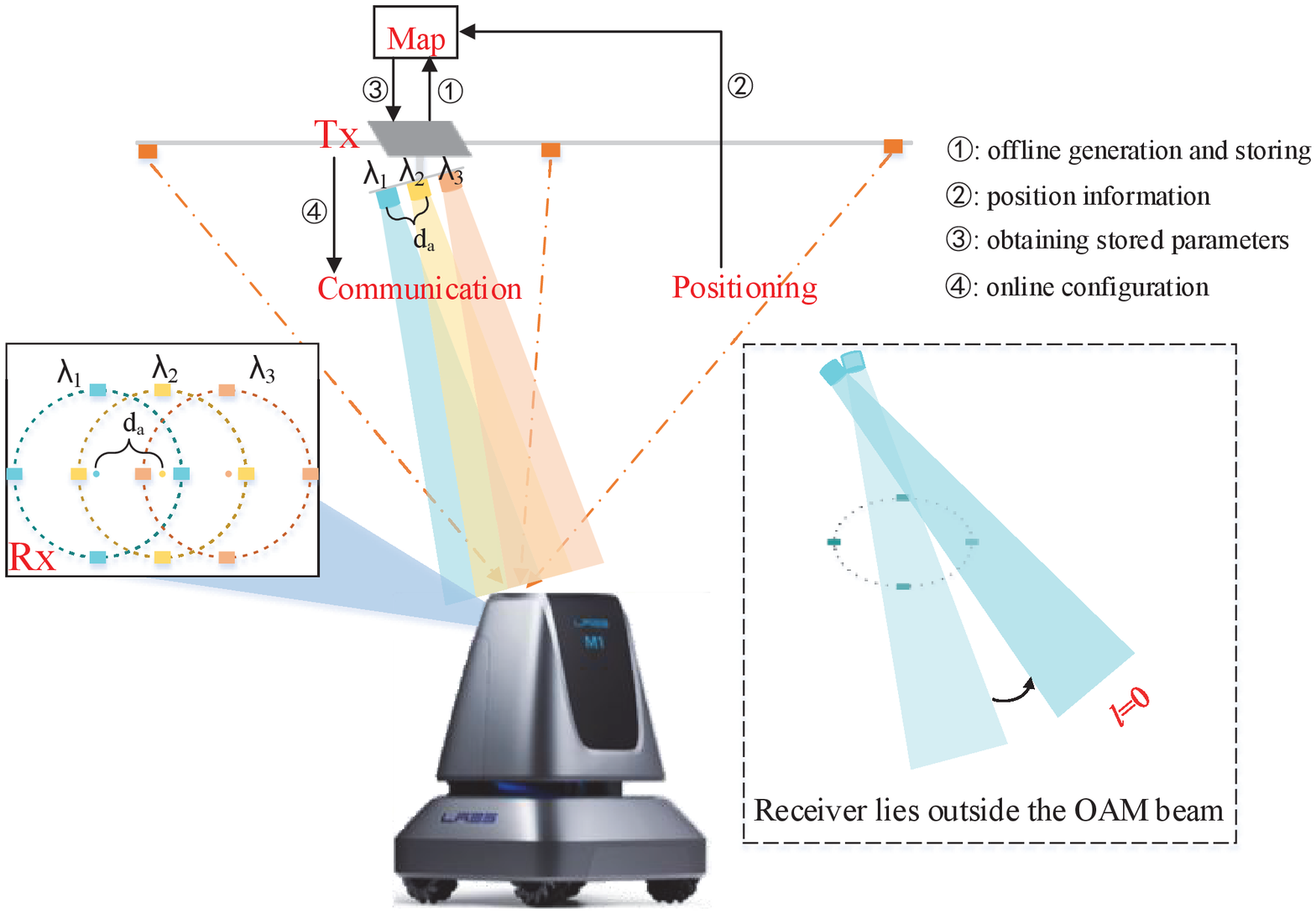}
	\caption{The scenario of the intelligent terminal equipped with mmWave WDM and OAM system.} \label{SysScene}
\end{figure*}

Consider a short-range LOS mmWave WDM system, which \textcolor{ColorRed}{is also equipped with} a positioning module, as shown in Figure \ref{SysScene}. The transmitter and receiver both adopt $I$ different carrier frequencies ($I =3$ in this example). Assume perfect band-pass filter on each carrier to remove the inter-carrier interference. Since the antenna of each carrier frequency can multiplex several OAM modes, the whole system can be regarded as consisting of $I$ parallel mmWave sub-systems. At the transmitter side, we utilize traveling-wave antennas or spiral phase plates (SPPs) to generate Laguerre-Gaussian (LG) beams with different OAM modes under coaxial condition~\cite{mmWaveOAM2014,ren2017line, 7968418, Travelingwave2015, 7014288}. The $I$ sub-systems form an uniform linear array with equal distance $d_a$. At the receiver side, we adopt uniform circular array (UCA) \cite{7797488, 8247288} for detection, where the interval between the centers of any two adjacent antennas is also $d_a$. \textcolor{ColorRed}{Benefiting} from the position information, the transmitting and receiving antennas are generally considered to be aligned \cite{ren2017line, zhang2016orbital}, which can also be realized by tracking schemes \cite{cai2019experimental, liu2019effective}. However, the receiver may lie outside the LG beam, which is shown in Figure \ref{SysScene}. In these regions, we assume that each sub-system aligns the beam axis with one of the receiving antennas in UCA. Since the whole system loses coaxial condition, we assume that only plane waves (i.e. the OAM mode number is $l=0$) are transmitted. 


We adopt LG method to characterize OAM beams\cite{7968418,8485721}. In cylindrical coordinate, the channel response on the $i$-th carrier with OAM mode $l$ at position $(r,\phi,z)$ is given by \cite{7968418, 8485721}
\begin{align} \label{equ.OAM_channel_response}
h_i^l(r,\phi,z) = &\frac{\sqrt{\zeta_i^l} \lambda_i}{4\pi d_{i,m}^l(z)} \left( \frac{r}{r_{i,max}^l(z)} \right)^{|l|} e^{\frac{\left( (r_{i,max}^l(z))^2 - r^2 \right) }{\omega_i^2(z)}}\cdot \notag \\
&e^{-j\frac{\pi(r^2 - (r_{i,max}^l(z))^2 ) }{\lambda_i R_i^l(z)}} e^{-j\frac{2\pi}{\lambda_i} d_{i,m}^l(z)} e^{-jl\phi}, 
\end{align} 
where $\phi$ is the transverse azimuthal angle \cite{Allen1992}; $\lambda_i$ is the carrier wavelength;  $\omega_i(z)=\omega_i\sqrt{1+(\frac{z}{z_R})^2} $ is the beam spot of the fundamental Gaussian beam and $\omega_i$ is the beam waist radius;  $z_R=\frac{\pi\omega_i^2}{\lambda_i}$ is the Rayleigh distance and assumed the same for all sub-systems; $R_i^l(z)=z\left[1 + \left( \frac{\pi\omega_i^2}{\lambda_i z}\right)^2\right]$ is the curvature radius;  $r_{i,max}^l(z) = \sqrt{\dfrac{|l|}{2}} \omega_i(z) = \omega_i \sqrt{\dfrac{|l|}{2} \left( 1 + \left (\dfrac{z}{z_R} \right )^2 \right) }$ is the radius of maximum energy strength region, in which the channel response follows Frris law \cite{7968418, 7160713}; $d_{i,m}^l(z) = \sqrt{(r_{i,max}^l(z))^2 + z^2}$; $e^{-jl\phi}$ denotes the helical phase distribution of OAM wave with mode $l$ ($l=0$ for a plane wave); $\zeta_i^l$ incorporates both antenna gain and system loss of $i$-th sub-system with OAM mode $l$ and is assumed to be 1. Therefore, the link gain can be written as
\begin{align} \label{equ.OAM_link_gain}
g_i^l(r,z) &= \left|h_i^l(r,z) \right|^2  \notag \\
	&= \frac{\lambda_i^{2}}{(4\pi)^{2}(d_{i,m}^l(z))^2} \left( \frac{r}{r_{i,max}^l(z)} \right)^{2|l|} e^{\frac{2\left( (r_{i,max}^l(z))^2 - r^2 \right) }{\omega_i^2(z)}}. 
\end{align} 

According to Equation \eqref{equ.OAM_channel_response}, the channel response can be represented as $h_i^l(r,\phi,z) = h_i^l(r,z) e^{-jl\phi} $. Due to inter-mode orthogonality, the OAM-based systems can coaxially transmit multiple OAM beams without inter-channel interference (ICI). We assume the same OAM modes employed in each sub-system, which can be perfectly separated at the receiver.


Figure \ref{SysScene} also presents the process of map-assisted method. The system solves optimization problems at all positions in an offline manner, classifies these results into a few patterns, and then stores these results into a map. After obtaining the position information, the system looks up the parameters from the map, and configures the transmitter and receiver by adopting these parameters. Note that, if the system lies in the environment with lots of small-scale fading, the channel gain cannot be determined even if the position of the transceiver is fixed. Therefore, the obtained map is not applicable in this scenario. In this paper, we mainly consider the mmWave frequency and short-range LOS environment, where the links are sparse and quasi-static. 
Apparently, quantization error will mismatch the position together with its optimal parameters, leading to performance loss. 
In the following section, we theoretically analyze the performance loss in the multi-dimensional constellation design, and find the features in maps via combination with numerical results.


\graphicspath{{figures/Fig_Constellation/}}

\section{ Multi-Dimensional Constellation Map for mmWave WDM systems }  \label{section.Cons_design}
In this section, we consider multi-dimensional constellation design and analyze the performance loss when adopting map-assisted approach.
\subsection{Multi-Dimensional Constellation Design} 
We assume that all OAM parallel sub-channels serve one user and jointly carry one multi-dimensional symbol. Let $\mathcal{L}$ denote the set of OAM modes, $L = |\mathcal{L}|$ denote the number of OAM modes in any sub-system, and $U = I L $ denote the total number of sub-channels. The sub-channel set is denoted by $ \{1, 2, \ldots, U \}$. The multi-dimensional constellation set is denoted by $\mathcal{C} = \{\mathbf{x}_1, \mathbf{x}_2, \ldots, \mathbf{x}_M\}$ with $\mathbf{x}_m \in \mathbb{C}^{U}$. We assume that the bandwidth of each sub-channel is the same.

At the receiver, the received signal can be written as $\mathbf{y} = \mathbf{H}\mathbf{x}+\mathbf{n}$, where $\mathbf{H} \in \mathbb{C}^{U \times U} $ is the parallel channel matrix assumed to be diagonal due to axis alignment and perfect band-pass filter in each sub-channel; $\mathbf{n} \in \mathbb{C}^{U}$ is the additive Gaussian noise vector with zeros mean and covariance matrix $N_0 \mathbf{I}$. We maximize the minimum Euclidean distance (MED) $d_{min}(\mathbf{H},\mathcal{C}) =\min_{\mathbf{x}_m, \mathbf{x}_n \in \mathcal{C}, m \ne n}\|\mathbf{H}(\mathbf{x}_m - \mathbf{x}_n)  \|_2$ for multi-dimensional constellation design \cite{7054450}.

Denoting $D \triangleq 2U$, we stack all symbol vectors of $\mathcal{C}$ to a single $MD \times 1$ column vector as
\begin{align}
\mathbf{x_p} = [\text{Re}\{\mathbf{x}_1\}^T, \text{Im}\{\mathbf{x}_1\}^T,  \cdots, \text{Re}\{\mathbf{x}_M\}^T, \text{Im}\{\mathbf{x}_M\}^T ]^T,
\end{align} 
where Re$\{\cdot\}$ and Im$\{\cdot\}$ represent the element-wise operations of extracting the real and imaginary components of a vector, respectively; $\{\cdot\}^T$ represents the transpose operation. The Euclidean distance of two symbols $ \mathbf{x}_m$ and $ \mathbf{x}_n$ can be rewritten as $\mathbf{x_p}^T \mathbf{E}_{mn} \mathbf{x_p}$, where 
\begin{small}
\begin{equation} \label{equ.E_ij}
\mathbf{E}_{mn} = \left[  
\begin{array}{ccccc}
\ddots &   &  & &  \\
& \mathbf{V_D} &  & -\mathbf{V_D} &  \\
&   & \ddots &  &  \\
& -\mathbf{V_D} & & \mathbf{V_D}&  \\
&   & &  &\ddots  \\
\end{array}
\right]_{MD \times MD},
\end{equation}
\end{small}
$\mathbf{V_D} \in \mathcal{R}^{D \times D} $ is given by
\begin{equation}\label{equ.V_D}
\mathbf{V}_{D} = \left[  
\begin{array}{cc}
\mathbf{H}^H\mathbf{H} & \mathbf{0} \\
\mathbf{0} &  \mathbf{H}^H\mathbf{H} \\
\end{array}
\right],
\end{equation}
$(\cdot)^H$ represents the conjugate transpose operation. In $\mathbf{E}_{mn}$, the $(m,m)$-th and $(n,n)$-th $D \times D$ blocks are $\mathbf{V_D}$; the $(m,n)$-th and $(n,m)$-th blocks are $-\mathbf{V_D}$; and other elements are all zero. The constellation optimization problem can be formulated as 
\begin{align} \label{equ.HighdimCons_original}
	\max_{\mathbf{x_p}} \quad & d_{ min} \notag\\
	\mbox{s.t.}\quad & \mathbf{x_p}^T \mathbf{E}_{mn} \mathbf{x_p} \geq d_{ min}^2, \quad \forall m,n, 1\leq m<n\leq M, \notag \\
	& \frac{1}{M}\left\| \mathbf{x_p}\right\|_2^2 \leq P_{sum},
\end{align}
where $P_{sum}$ denotes the total transmit power limit and is set as $P_{sum}=1 $ in this paper. The constraint $\mathbf{x_p}^T \mathbf{E}_{mn} \mathbf{x_p} \geq d_{ min}^2$ is non-convex and can be approximately transformed into the following linear constraint:
\begin{align} \label{equ.convexC}
\mathbf{x_p}^T \mathbf{E}_{mn} \mathbf{x_p} & \approx 2 \cdot \mathbf{x_p}_{(k-1)}^T \mathbf{E}_{mn} \mathbf{x_p}_{(k)} - 
\mathbf{x_p}_{(k-1)}^T \mathbf{E}_{mn} \mathbf{x_p}_{(k-1)}  \notag \\
&\geq d_{ min}^2, \quad \forall m,n, 1\leq m<n\leq M.
\end{align}

Based on Equation \eqref{equ.convexC}, the optimization problem \eqref{equ.HighdimCons_original} can be solved via an iterative procedure (subscript $k$ represents the $k$-th iteration), which is generally computationally intensive due to the high optimization dimension, especially when $U$ or $M$ is large. For example, when $I=2$, $L=2$ and $M=64$, the optimization dimension of $\mathbf{x_p}$ is $2M \times L \times I=512$, resulting in the difficulty of real-time optimization. A multi-dimensional constellation map can be constructed as a look-up table to reduce the real-time computational complexity. 


\subsection{Multi-Dimensional Constellation Map  }
We characterize the constellation design according to the receiver location. We divide the space into regions and calculate the optimal power allocation for each region, assuming that the optimal power allocation for each region center can represent that for this region.


\begin{figure}[htbp]
	\centering
	\includegraphics[width=3.5in]{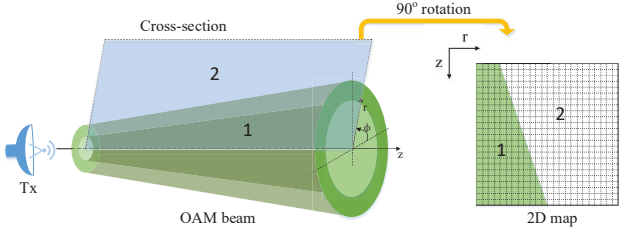}
	\caption{The OAM transmission in the 3D space, and the arbitrary cross-section.} \label{2Dmap}
\end{figure}

In general, the receiver moves in a three-dimensional (3D) space. Since the link gain is independent of azimuth angle $\phi$, the 3D map can be reduced due to rotational symmetry, and we can only consider vertical distance $z$ along the axis and horizontal distance $r$ from the axis. As shown in Figure \ref{2Dmap}, the two-dimensional (2D) map is an arbitrary cross-section of the OAM beam. The 2D map is divided into two regions, where region 1 is covered by the OAM beam and region 2 lies outside the OAM beam. The power allocation map can be generated according to the 2D map and provide prior knowledge to configure the transceiver.

To reduce the demand of look-up table storing and searching in map-assisted method, we quantify the region of interest into a small set of positions. Then, the system uses the optimal solution of one position as that of some other positions, and all positions in the map will be classified into several categories. More specifically, we adopt normalized MED difference as the criterion when generating the multi-dimensional constellation map, where the multi-dimensional constellation for a position in a certain region is adopted to represent that for all positions in this region. The normalized MED difference for position $q_1$ using the constellation for position $q_2$ can be expressed as  
\begin{align}  \label{equ.MED_distortion}
		\Delta \bar{d}_{min}(q_1, q_2) = \left| 1 - \frac{d_{min}(\mathbf{H}^{q_2},\mathcal{C}^{(q_1)} )}{d_{min}(\mathbf{H}^{q_2},\mathcal{C}^{(q_2)} )}  \right| ,
\end{align}
where $\mathbf{H}^{q_2}$ is the channel matrix at position $q_2$; $\mathcal{C}^{(q_1)}$ and $\mathcal{C}^{(q_2)}$ are the optimal multi-dimensional constellations for positions $q_1$ and $q_2$, respectively.

We adopt clustering approach to quantify the parameters of all positions with a distortion threshold $\tau_{d_{min}}$. Letting $\mathcal{Q}$ denote the set of all positions, we randomly select a position $q_a$ from $\mathcal{Q}$ with the optimal multi-dimensional constellation $\mathcal{C}^{(q_a)}$. For other positions, if the normalized MED differences from $\mathcal{C}^{(q_a)}$ is lower than threshold $\tau_{d_{min}}$, we classify these positions and $q_a$ into one category and remove these positions from $\mathcal{Q}$. Moreover, the constellation of this category is set to be $\mathcal{C}^{(q_a)}$. The random position selection and classification continues until all positions are classified into certain categories. Then, we store the generated multi-dimensional constellation map and calculate the total normalized MED difference $\Delta \bar{d}_s$, which is the summation of the normalized MED differences at all positions. We repeat the whole above process $C_d$ times and select the map with the lowest $\Delta \bar{d}_s$ as the final multi-dimensional constellation map $\mathcal{C}_{map}$. In this paper, we set $\tau_{d_{min}} = 0.15$ and $C_d = 100$.

\subsection{MED-difference Analysis on Multi-Dimensional Constellation Map  } \label{subsection.Robustness_high_dim_cons}

In this subsection, we investigate the feature of MED-difference, which can further determine the feature of the multi-dimensional constellation map. 

We first consider the influence of channel condition. To facilitate the analysis, we consider link gain $g_i^l(r,z)$ rather than channel response $h_i^l(r,\phi,z)$, and focus on the link gain ratios within OAM beam regions, where $\mathcal{L} $ consists of several OAM modes. Denote $\lambda_a$ as an arbitrary carrier wavelength. Let $r = \beta r_{a,max}^{l_m}(z)$, where $\beta$ is a non-negative number and $l_m$ is an arbitrary nonzero OAM mode. It is seen that when $\lambda_a$ and $l_m$ are fixed, radius $r$ can be replaced by $\beta$ to determine the horizontal distance from the beam axis.  Consider two sub-channels with different wavelengths $\lambda_i$ and  $\lambda_j, 1\leq i, j \leq I$, and the same OAM mode $l\in \mathcal{L}$. Denoting $a_{i,j}^{l}(\beta, z)  \triangleq \frac{g_i^l(\beta r_{a,max}^{l_m}(z), z)}{g_j^l(\beta r_{a,max}^{l_m}(z), z) }$ for $\lambda_i>\lambda_j$, we have
\begin{align} \label{equ.gi_gj}
	&a_{i,j}^{l}(\beta, z) \notag \\ 
	&=  \left( \frac{\lambda_j }{\lambda_i} \right)^{|l|-2} \frac{ \omega_j^2 \frac{|l|}{2} \left( 1+\frac{z^2 }{z_R^2} \right) + z^2 }{ \omega_i^2 \frac{|l|}{2} \left( 1+\frac{z^2 }{z_R^2} \right) + z^2} e^{ \beta^2 |l_m|\left( \frac{\lambda_i}{\lambda_j} - 1 \right)} .
\end{align}
Note that term $\frac{ \omega_j^2 \frac{|l|}{2} \left( 1+\frac{z^2 }{z_R^2} \right) + z^2 }{ \omega_i^2 \frac{|l|}{2} \left( 1+\frac{z^2 }{z_R^2} \right) + z^2}$ increases monotonically with $z$ and converges to $1$ when $\lambda_i>\lambda_j$. Moreover, this term converges rapidly because $\omega^2 = z_R \lambda /\pi \ll 1$ in short-range mmWave systems. For instance, considering that $l=+2$, $z_R = 4$ m and two carrier frequencies $f_i = 60$ GHz and $f_j = 65$ GHz, this term is larger than 0.995 for $z > 0.1$ m. Hence, the above link gain ratio can be approximated as $a_{i,j}^{l}(\beta, z) \approx   \left( \frac{\lambda_j }{\lambda_i} \right)^{|l|-2} e^{ \beta^2 |l_m|\left( \frac{\lambda_i}{\lambda_j} - 1 \right)}$.  Then, consider two sub-channels with different OAM modes $l_1, l_2  \in \mathcal{L}$, and the same wavelength $\lambda_i, 1\leq i  \leq I$. Assume $|l_1|$ and $|l_2|$ are small integers (e.g., $l_1 = +2$ and $l_2 = +1$). Denoting $a_i^{l_1,l_2}(\beta, z)  \triangleq \frac{g_i^{l_1}(\beta r_{a,max}^{l_m}(z), z)}{g_i^{l_2}(\beta r_{a,max}^{l_m}(z), z)}$ for $|l_1|>|l_2|$, we have  
\begin{align} \label{equ.gl1_gl2}
	a_i^{l_1,l_2}(\beta, z) & \approx  \left( \frac{\lambda_a }{\lambda_i} \right)^{|l_1|-|l_2|}  \left( \beta^2 |l_m| \right)^{|l_1|-|l_2|} \frac{|l_2|^{|l_2|}}{|l_1|^{|l_1|}} e^{ |l_1|-|l_2|}.
\end{align}

Considering link gain ratios at different positions, we have
\begin{align} \label{equ.aij1_aij2}
	&\frac{a_{i,j}^{l,(1)}}{a_{i,j}^{l,(2)}} =\frac{a_{i,j}^{l}(\beta_1, z_1) }{a_{i,j}^{l}(\beta_2, z_2)} = e^{  |l_m|\left( \frac{\lambda_i}{\lambda_j} - 1 \right) (\beta_1^2 - \beta_2^2) } , \notag \\
	& \frac{a_i^{l_1,l_2,(1)}}{a_i^{l_1,l_2,(2)}} =\frac{a_i^{l_1,l_2}(\beta_1, z_1)}{a_i^{l_1,l_2}(\beta_2, z_2)} = \left( \frac{\beta_1}{\beta_2} \right) ^{2(|l_1| - |l_2|)}.
\end{align}

According to Equation \eqref{equ.aij1_aij2}, the link gain ratios are primarily determined by $\beta_1$ and $\beta_2$, and are both equal to 1 if $\beta_1 = \beta_2$. This indicates that for the positions with the same $\beta$, $a_{i,j}^{l}$ and $a_i^{l_1,l_2}$ remain unchanged, i.e., all sub-channels have uniform link gain ratio. More specifically, for any $i$ and $l_1$ ($1\leq i \leq I, l_1\in \mathcal{L}$) and fixed $j$ and $l_2$ ($1\leq j \leq I, l_2\in \mathcal{L}$), we have $g_i^{l_1} = a_{i,j}^{l_1} g_j^{l_1} = a_{i,j}^{l_1} a_{j}^{l_1,l_2} g_j^{l_2} $. For two positions 1 and 2 with the same $\beta$, since $a_{i,j}^{l_1}$ and $a_{j}^{l_1,l_2}$ are unchanged, we have $g_i^{l_1,(1)} = a_{i,j}^{l_1} a_{j}^{l_1,l_2} g_j^{l_2,(1)} $ and $g_i^{l_1,(2)} = a_{i,j}^{l_1} a_{j}^{l_1,l_2} g_j^{l_2,(2)} $. Letting $g_j^{l_2,(2)} = \alpha^2 g_j^{l_2,(1)}$, where $\alpha$ is a positive number, we have $g_i^{l_1,(2)} = \alpha^2 g_i^{l_1,(1)}$ for any $i$ and $l_1$, $1\leq i \leq I, l_1\in \mathcal{L}$. Denoting link gain matrix $\mathbf{G} = \text{diag}\left(  g_1^{l_1}, \cdots, g_1^{l_L}, \cdots, g_I^{l_1}, \cdots, g_I^{l_L}\right) $ with $\text{diag}(\cdot)$ being the diagonalization operation, we have $\mathbf{G}^{(2)} = \alpha^2 \mathbf{G}^{(1)}$.

Since $\mathbf{G} = \mathbf{H}^H \mathbf{H}$, for any two positions satisfying uniform link gain ratio $\mathbf{G}^{(2)} = \alpha^2 \mathbf{G}^{(1)}$, we have ${\mathbf{H}^{(2)}}^H \mathbf{H}^{(2)} = \alpha^2 {\mathbf{H}^{(1)}}^H \mathbf{H}^{(1)} $. Combining Equations \eqref{equ.E_ij} and \eqref{equ.V_D}, the first constraint under $\mathbf{H}^{(2)}$ can be expressed as
\begin{align} 
	\mathbf{x_p}^T \mathbf{E}_{mn}^{(2)} \mathbf{x_p} = \alpha^2 \cdot \mathbf{x_p}^T &\mathbf{E}_{mn}^{(1)} \mathbf{x_p} \geq d_{ min}^2, \notag \\
	& \quad \forall m,n, 1\leq m<n\leq M .
\end{align}
It is easy to see that the original optimization problem \eqref{equ.HighdimCons_original} under $\mathbf{H}^{(2)}$ is equivalent to that under $\mathbf{H}^{(1)}$, with only scaling difference. 

For OAM beam region, the variations of the link gains in different sub-channel are proportional only at the positions with the same $\beta$, and are not proportional at other positions. Hence, we consider the influence of this non-proportional channel variation. Assume that ${\mathbf{H}^{(2)}}^H \mathbf{H}^{(2)} = \left( \alpha \mathbf{H}^{(1)} + \Delta \mathbf{H} \right)^H \left( \alpha \mathbf{H}^{(1)} + \Delta \mathbf{H} \right)  $, where $\Delta \mathbf{H}$ stands for the deviation. Let $\mathcal{C}^{(1)}$ and $\mathcal{C}^{(2)}$ denote the optimal multi-dimensional constellation set for $\alpha \mathbf{H}^{(1)}$ and $ \mathbf{H}^{(2)}$, respectively. Assume that $\{\mathbf{x}_1^{(1)}, \tilde{\mathbf{x}}_1^{(1)}\}$ and $\{\mathbf{x}_1^{(2)}, \tilde{\mathbf{x}}_1^{(2)}\}$ are the corresponding MED constellation pairs for $\alpha \mathbf{H}^{(1)}$ using $\mathcal{C}^{(1)}$ and $\mathcal{C}^{(2)}$, respectively; $\{\mathbf{x}_2^{(1)}, \tilde{\mathbf{x}}_2^{(1)}\}$ and $\{\mathbf{x}_2^{(2)}, \tilde{\mathbf{x}}_2^{(2)}\}$ are the corresponding MED constellation pairs for $\mathbf{H}^{(2)}$ using $\mathcal{C}^{(1)}$ and $\mathcal{C}^{(2)}$, respectively. Then, we have the following result on the normalized MED difference.
\begin{theorem} \label{theorem.dmin_diff_nonuniform_linkgain}
	Given $\mathbf{H}^{(1)}$ and $\mathbf{H}^{(2)}$ that satisfy ${\mathbf{H}^{(2)}}^H \mathbf{H}^{(2)} = \left( \alpha \mathbf{H}^{(1)} + \Delta \mathbf{H} \right)^H \left( \alpha \mathbf{H}^{(1)} + \Delta \mathbf{H} \right)  $, we have the following upper bound
	\begin{align}
	\Delta \bar{d}_{min}^{\Delta \mathbf{H}} &\triangleq  1 - \frac{d_{min}(\mathbf{H}^{(2)}, \mathcal{C}^{(1)} ) }{d_{min}(\mathbf{H}^{(2)}, \mathcal{C}^{(2)} )} \notag \\
	&\leq \frac{ \|\Delta \mathbf{H} \|_F(\| \mathbf{x}_1^{(2)}-\tilde{\mathbf{x}}_1^{(2)}\|_2 + \|  \mathbf{x}_2^{(1)}-\tilde{\mathbf{x}}_2^{(1)}\|_2)}{\|\mathbf{H}^{(2)} (\mathbf{x}_2^{(2)}-\tilde{\mathbf{x}}_2^{(2)})\|_2} .
	\end{align}	
	Generally, term $(\|\mathbf{x}_1^{(2)}-\tilde{\mathbf{x}}_1^{(2)}\|_2 + \| \mathbf{x}_2^{(1)}-\tilde{\mathbf{x}}_2^{(1)}\|_2) / \|\mathbf{H}^{(2)} (\mathbf{x}_2^{(2)}-\tilde{\mathbf{x}}_2^{(2)})\|_2$ is finite, we have $\Delta \bar{d}_{min}^{\Delta \mathbf{H}} =  O(\|\Delta \mathbf{H} \|_F)$.
	
	\begin{proof}
		See Appendix \ref{appendix.dmin_diff_nonuniform}.
	\end{proof}
\end{theorem}
Since $d_{min}(\mathbf{H}^{(2)}, \mathcal{C}^{(2)} ) \geq d_{min}(\mathbf{H}^{(2)},  \mathcal{C}^{(1)} )$ always holds, we have $\Delta \bar{d}_{min}^{\Delta \mathbf{H}} \geq 0$. Theorem \ref{theorem.dmin_diff_nonuniform_linkgain} indicates that the order of the asymptotic convergence rate of $\Delta \bar{d}_{min}^{\Delta \mathbf{H}}$ with respect to $\Delta \mathbf{H}$ is not larger than that of $\|\Delta \mathbf{H} \|_F$. Hence, $\Delta \bar{d}_{min}^{\Delta \mathbf{H}}$ can be sufficiently small if $\|\Delta \mathbf{H} \|_F$ is sufficiently small, and the systems under $\mathbf{H}^{(1)}$ and $\mathbf{H}^{(2)}$ can adopt a fixed multi-dimensional constellation.

Since it is difficult to obtain an accurate theoretical position variation range, we attempt to qualitatively analyze the classification. Substituting $r = \beta r_{a,max}^{l_m}(z)$ and $r_{i,max}^l(z) = \beta^l_{i,max} r_{a,max}^{l_m}(z) = \sqrt{\frac{\lambda_i |l|}{\lambda_a |l_m|} } r_{a,max}^{l_m}(z)$ into Equation \eqref{equ.OAM_link_gain}, the link gain can be rewritten as
\begin{align} \label{equ.OAM_link_gain_new}
	& g_i^l(\beta r_{a,max}^{l_m}(z),z) \notag \\
	&= \frac{\zeta_i^l \lambda_i^{2}}{(4\pi)^{2}(d_{i,m}^l(z))^2} \left( \frac{\lambda_a |l_m|}{\lambda_i |l|} \right)^{|l|} \beta^{2|l|}e^{|l|}e^{-\frac{\lambda_a |l_m|}{\lambda_i} \beta^2}.
\end{align} 
Since $\beta^l_{i,max} $ can represent the position with the maximum link gain for any $z$, we have that $g_i^l(\beta r_{a,max}^{l_m}(z),z)$ first increases as $\beta$ increases within range $0 < \beta < \sqrt{\frac{\lambda_i |l|}{\lambda_a |l_m|}   }$, and then decreases as $\beta$ increases within range $ \beta > \sqrt{\frac{\lambda_i |l|}{\lambda_a |l_m|} }$. Specially, for $l=0$, $g_i^l(\beta r_{a,max}^{l_m}(z),z)$ always decreases as $\beta$ increases. For any two OAM modes $l_1$ and $l_2$ satisfying $|l_1|>|l_2|$, $a_i^{l_1,l_2}(\beta, z) $ increases as $\beta$ increases according to Equation \eqref{equ.gl1_gl2}. However, when both $|l_1|$ and $|l_2|$ are small and $|l_1|$ is close to $|l_2|$, there exist some positions where $a_i^{l_1,l_2}(\beta, z) $ is a moderate value. For instance, for $|l_1|=1$, $|l_2|=0$ and $\zeta_i^{l_1}=\zeta_i^{l_2}$, $a_i^{l_1,l_2}(\beta^{l_1}_{i,max}, z) = e$, which implies that $a_i^{l_1,l_2}(\beta, z) $ is moderate around $\beta^{l_1}_{i,max}$. For short-range LOS links, $g_i^{l_1}(\beta r_{a,max}^{l_m}(z),z)$ is high within this range, and thus $g_i^{l_2}(\beta r_{a,max}^{l_m}(z),z)$ is also high due to moderate $a_i^{l_1,l_2}(\beta, z) $. To increase the MED of the multi-dimensional constellation at these positions, the constellation symbols will disperse more evenly on all sub-channels, especially when the modulation order is large. Hence, the location of constellation points is close to the evenly spaced distribution. We consider these regions as the OAM beam regions with high signal strength. 

In the regions where $\beta$ is far away from $\beta^l_{i,max}$, $g_i^l(\beta r_{a,max}^{l_m}(z),z)$ in the sub-channel with indexes $i$ and $l$ is low, and the discrepancy between any two sub-channels is large. Thus, constellation symbols will be more likely to gather on sub-channels with high link gain. We consider these regions as the boundary regions (left or right region) of the OAM beam with mode $l$ in the $i$-th sub-system. For any $l\neq 0$, assuming $0 < \Delta \beta < \beta^l_{i,max}$, we have
\begin{align} \label{equ.OAM_link_gain_deltabeta}
	a_r &\triangleq \frac{g_i^l(\beta^l_{i,max} + \Delta \beta,z)}{g_i^l(\beta^l_{i,max} - \Delta \beta,z)}\notag \\
	& = \left( 1 + \frac{2 \Delta \beta }{\beta^l_{i,max} -\Delta \beta }\right)^{2|l|} e^{\frac{|l_m| \lambda_a}{\lambda_i} 4\Delta \beta \beta^l_{i,max}} . 
\end{align} 
Since $\partial a_r / \partial \Delta \beta > 0$ always holds, we have $a_r > a_r|_{\Delta \beta = 0} = 1$. This implies that the signal strength in the left boundary regions of OAM beams decays more rapidly than that in right boundary regions as $\Delta \beta$ increases. More specifically, compared with range $\beta > \beta^l_{i,max}$, the channel matrix $\mathbf{H}$ changes more rapidly within range $0 < \beta < \beta^l_{i,max} $ and thus $\|\Delta \mathbf{H} \|_F$ is larger when given a same $\Delta \beta$. Hence, there will be multiple categories in the left region of $\beta^l_{i,max} = \sqrt{\frac{\lambda_i |l|}{\lambda_a |l_m|} }$ in the multi-dimensional constellation map. Compared with the right region of $\beta^l_{i,max}$, the areas of these categories will be smaller.

Then, we also consider the power pre-allocation system, where the power in each sub-channel is configured only when the communication link is established, but not convenient for real-time adjustment. This situation will occur when the system adopts multiple separate transmitters, such as SPPs, to generate different OAM beams \cite{mmWaveOAM2014}. 
Once the communication link is established, the system can be considered as adopting fixed-power allocation. Letting $\mathbf{p}\triangleq[P_1, P_2, \ldots, P_U]^T$ denote the allocated power vector over all sub-channels and $\mathbf{A} \triangleq \sqrt{\text{diag}(\mathbf{p})}$, the multi-dimensional constellation symbol $\mathbf{x} \in \mathcal{C}$ satisfying $\mathbf{x}= \mathbf{A} \mathbf{s}$, where $\mathbf{s}$ is the multi-dimensional constellation symbol removing power matrix $\mathbf{A}$. Let $\mathcal{S} \triangleq \{\mathbf{s}_1, \mathbf{s}_2, \ldots, \mathbf{s}_M\}$ denote the set of $\mathbf{s}$. The MED can be rewritten as $d_{min} (\mathbf{H}, \mathbf{p}, \mathcal{S}) = \|\mathbf{H}\mathbf{A}(\mathbf{s}_m - \mathbf{s}_n) \|_2 $, where $\mathbf{s}_m$ and $\mathbf{s}_n$ are the MED constellation pair of $\mathcal{S}$. Moreover, each sub-channel has individual power constraint, and the second constraint of the optimization problem \eqref{equ.HighdimCons_original} can be replaced by $\frac{1}{M}\left\| \mathbf{J}_n \mathbf{x_p}\right\|_2^2 \leq P_n, \forall n \in \left\lbrace 1,2, \cdots, U\right\rbrace $, where $\mathbf{J}_n = \text{diag}( \mathbf{J}(n, :))$ and $\mathbf{J}$ is the $MD \times MD$ sub-channel selection matrix with only 0 and 1 elements. Since the first constraint remains unchanged, Theorem \ref{theorem.dmin_diff_nonuniform_linkgain} still holds in this case.

Compared with the system with total power constraint system, since the power in each sub-channel is difficult to change, the system with fixed-power vector may suffer a large performance loss. However, we can confirm that this performance loss can be small at some positions where the difference between the fixed-power vector and the optimal power vector is small. Assume that $\mathbf{p}^{(o)}$ and $\mathbf{p}^{(f)}$ are the optimal and fixed-power allocation vectors, respectively. $\mathbf{p}^{(o)}$ can be extracted from the optimal multi-dimensional constellation $\mathcal{C}^{(o)}$. Assume $\mathcal{S}^{(o)}$ and $\mathcal{S}^{(f)}$ are the optimal constellation sets for $\mathbf{H}$ under $\mathbf{p}^{(o)}$ and $\mathbf{p}^{(f)}$, respectively. Assume that $\{\mathbf{s}_o^{(o)}, \tilde{\mathbf{s}}_o^{(o)} \}$ and $\{\mathbf{s}_o^{(f)}, \tilde{\mathbf{s}}_o^{(f)}\}$ are the corresponding MED constellation pairs for $\mathbf{p}^{(o)}$ using $\mathcal{S}^{(o)}$ and $\mathcal{S}^{(f)}$, respectively; $\{\mathbf{s}_f^{(o)}, \tilde{\mathbf{s}}_f^{(o)} \}$ and $\{\mathbf{s}_f^{(f)}, \tilde{\mathbf{s}}_f^{(f)}\}$ are the corresponding MED constellation pairs for $\mathbf{p}^{(f)}$ using $\mathcal{S}^{(o)}$ and $\mathcal{S}^{(f)}$, respectively. We have the following result on the normalized MED difference.



\begin{theorem} \label{theorem.dmin_diff_power_allocation}
	Letting $\mathcal{S}_d$ denote the set of $ \left\{ \mathbf{s}_o^{(o)}-\tilde{\mathbf{s}}_o^{(o)}, \mathbf{s}_o^{(f)}-\tilde{\mathbf{s}}_o^{(f)}, \mathbf{s}_f^{(o)}-\tilde{\mathbf{s}}_f^{(o)},  \mathbf{s}_f^{(f)}-\tilde{\mathbf{s}}_f^{(f)} \right\}$ and $\mathbf{A}^{(o)}=\sqrt{\text{diag}(\mathbf{p}^{(o)})}$, we have the following upper bound
	\begin{align}
		\Delta \bar{d}_{min}^{\mathbf{p}} &\triangleq \left| 1 - \frac{d_{min}(\mathbf{H}, \mathbf{p}^{(f)}, \mathcal{S}^{(f)} ) }{d_{min}(\mathbf{H}, \mathbf{p}^{(o)}, \mathcal{S}^{(o)} )}  \right| \notag \\
		&\leq \left\| \mathbf{p}^{(o)} - \mathbf{p}^{(f)} \right\|_2 \cdot \max_{\mathbf{s}_d \in \mathcal{S}_d} \left\{ \frac{  \|\mathbf{H}\mathbf{s}_d\|_2^2}{\|\mathbf{H}\mathbf{A}^{(o)}\mathbf{s}_d\|_2^2}\right\}.
	\end{align}
	Generally, term $\max_{\mathbf{s}_d \in \mathcal{S}_d} \left\{    \|\mathbf{H}\mathbf{s}_d\|_2^2 / \|\mathbf{H}\mathbf{A}^{(o)}\mathbf{s}_d\|_2^2\right\}$ is finite, we have $\Delta \bar{d}_{min}^{\mathbf{p}} = O(\left\| \mathbf{p}^{(o)} - \mathbf{p}^{(f)} \right\|_2)$.
	
	\begin{proof}
		See Appendix \ref{appendix.dmin_diff}.
	\end{proof}
\end{theorem}

Based on Theorem \ref{theorem.dmin_diff_power_allocation}, the order of the asymptotic convergence rate of $\Delta \bar{d}_{min}^{\mathbf{p}}$ with respect to $\mathbf{p}^{(f)}$ is not larger than the order of $\left\| \mathbf{p}^{(o)} - \mathbf{p}^{(f)} \right\|_2$. This indicates that compared with the method using $\mathbf{p}^{(o)}$ to generate multi-dimensional constellations, the performance loss of using $\mathbf{p}^{(f)}$ can be sufficiently small if $\left\| \mathbf{p}^{(o)} - \mathbf{p}^{(f)} \right\|_2$ is sufficiently small. Hence, to avoid large performance loss of the system with fixed-power allocation, we can set $\mathbf{p}^{(f)}$ close to $\mathbf{p}^{(o)}$. When the power vector is fixed, we can adopt $\mathcal{S}^{(f)}$ to generate the multi-dimensional constellation map.


\section{Numerical Results for Multi-Dimensional Constellation Map}\label{section.NumericalResults}
In this section, numerical results are presented to verify the above theoretical analysis.

\begin{figure}[htbp]
	\centering
	\subfloat[The difference between $\mathbf{p}^{(f)}$ and $\mathbf{p}^{(o)}$.]{
		\includegraphics[width=2.7 in] {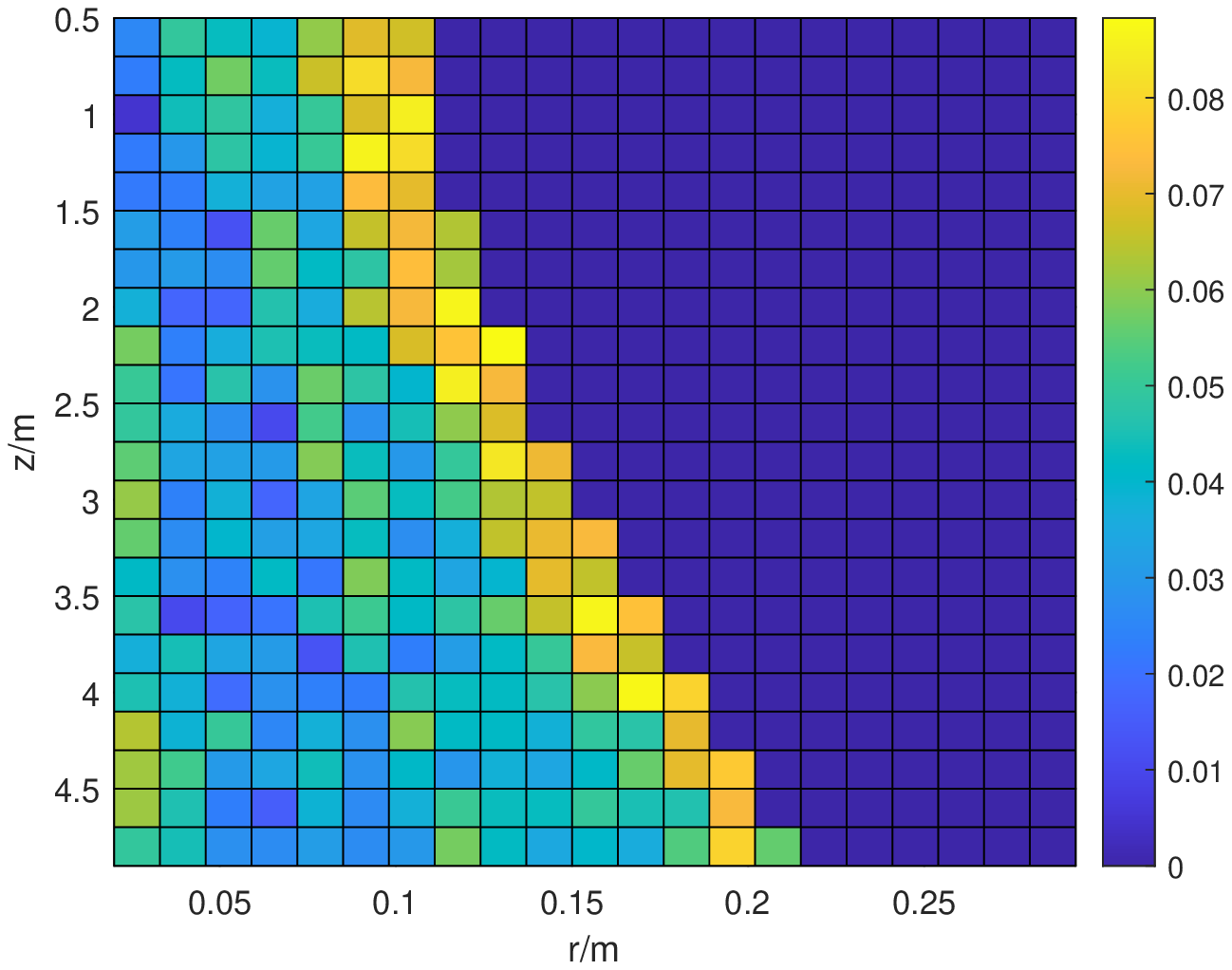}\label{diff_op_ep_01.a}
	}

	\subfloat[The normalized MED difference $\Delta \bar{d}_{min}^{\mathbf{p}}$.]{
		\includegraphics[width=2.7 in] {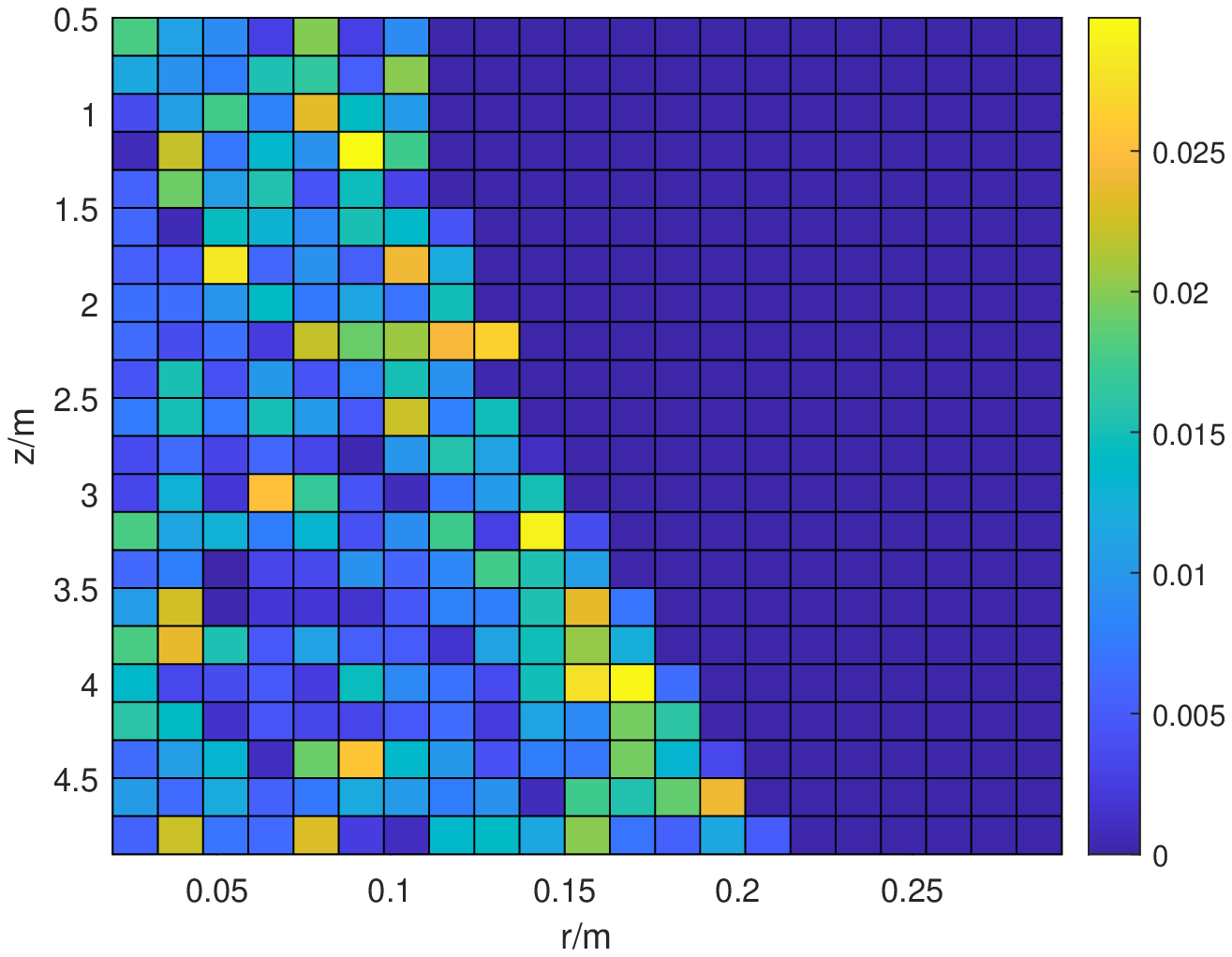}\label{diff_op_ep_01.b}
	}	
	\caption{The performance loss of the system with equal power allocation and $\mathcal{L}=\{0,+1\}$. }\label{diff_op_ep_01}
\end{figure}

\begin{figure}[htbp]
	\centering
	\subfloat[The difference between $\mathbf{p}^{(f)}$ and $\mathbf{p}^{(o)}$.]{
		\includegraphics[width=2.7 in] {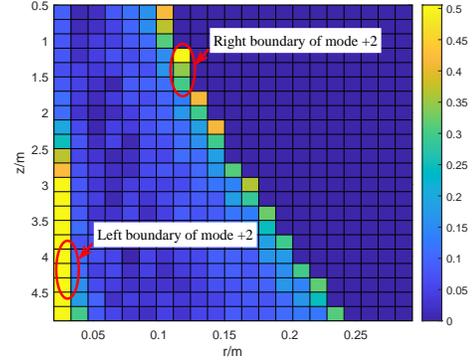}\label{diff_op_ep_02.a}
	}
	
	\subfloat[The normalized MED difference $\Delta \bar{d}_{min}^{\mathbf{p}}$.]{
		\includegraphics[width=2.7 in] {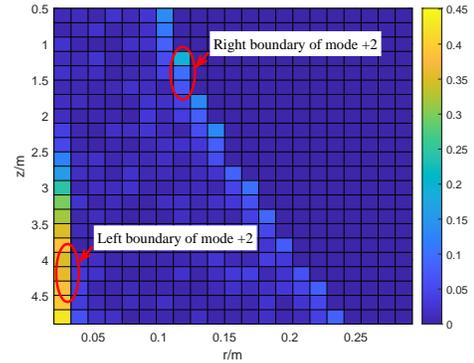}\label{diff_op_ep_02.b}
	}	
	\caption{The performance loss of the system with equal power allocation and $\mathcal{L}=\{0,+2\}$. }\label{diff_op_ep_02}
\end{figure}

We first consider the system with equal power allocation. Assume that the noise power is $N_0=10^{-10}$ and the number of sub-systems is $I=2$. The frequency of the first sub-system is $f_1 = c/\lambda_1 = 60$ GHz, where $c$ is the speed of light. The frequency interval is $ \triangle f=1$ GHz and the number of multi-dimensional symbols $M=64$. Consider the system with $\mathcal{L}=\{0,+1\}$. Figure \ref{diff_op_ep_01.a} shows the power allocation difference $\left\| \mathbf{p}^{(o)} - \mathbf{p}^{(f)} \right\|_2$. It is seen that the difference is always small in the OAM beam region. The reason is that in the short-range LOS environment, the SNRs of all sub-channels are high, generally larger than 15dB when $N_0=10^{-10}$. Hence, the constellation symbols in $\mathcal{C}^{(o)}$ disperse more evenly on all sub-channels, and power vector $\mathbf{p}^{(o)}$ extracted from $\mathcal{C}^{(o)}$ is close to equal power allocation. Figure \ref{diff_op_ep_01.b} shows the normalized MED difference $\Delta \bar{d}_{min}^{\mathbf{p}}$. We can observe that $\Delta \bar{d}_{min}^{\mathbf{p}}$ is always small, which is consistent with the feature in Theorem \ref{theorem.dmin_diff_power_allocation}. Considering the system with $\mathcal{L}=\{0,+2\}$, we can see from Figure \ref{diff_op_ep_02} that the power allocation difference and the normalized MED difference are also small in most positions. However, these differences will be very large in the boundary regions of the OAM beam with mode +2. The reason is that compared to the beam with mode +1, the beam with mode +2 has a larger divergence. In these boundary regions, the SNRs of the sub-channels with mode +2 are very low, and the constellation symbols in $\mathcal{C}^{(o)}$ will gather on the sub-channels with mode 0. Thus, if equal power allocation scheme is still adopted, the system will suffer a large performance loss.  

According to the above results, the fixed-power system in non-boundary region can adopt equal power allocation with a negligible performance loss. While in boundary region, equal power allocation among all sub-channels is not proper, and proper power pre-allocation scheme is required. Since the MED only depends on $\mathcal{S}^{(f)}$ at a certain position, we can adopt $\mathcal{S}^{(f)}$ rather than $\mathcal{C}^{(o)}$ to generate the multi-dimensional constellation map, which has the same features as that in the system with total power constraint.

\begin{figure}[htbp]
	\centering
	\subfloat[$\mathcal{L}=\{0,+1\}$, $\triangle f=1$ GHz, $M=64$]{
		\includegraphics[width=2.7 in] {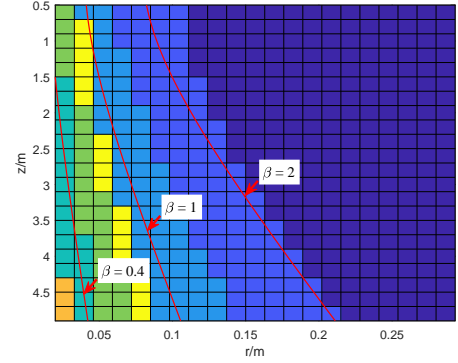}\label{dmin_map.a}
	}

	\subfloat[$\mathcal{L}=\{0,+2\}$, $\triangle f=1$ GHz, $M=64$]{
		\includegraphics[width=2.7 in] {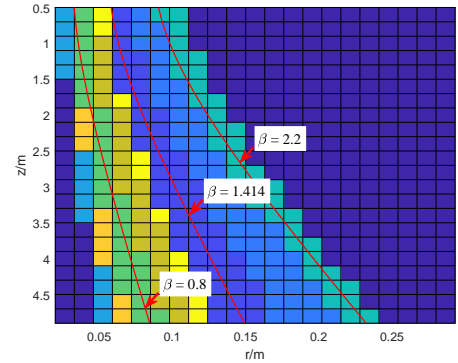}\label{dmin_map.b}
	}

	\subfloat[$\mathcal{L}=\{0,+1\}$, $\triangle f=5$ GHz, $M=64$]{
	\includegraphics[width=2.7 in]
	 {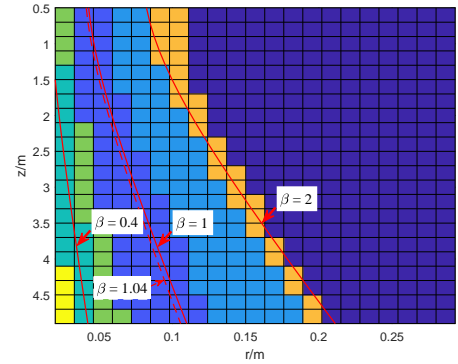}\label{dmin_map.c}
    }

	\subfloat[$\mathcal{L}=\{0,+1\}$, $\triangle f=1$ GHz, $M=32$]{
	\includegraphics[width=2.7 in] {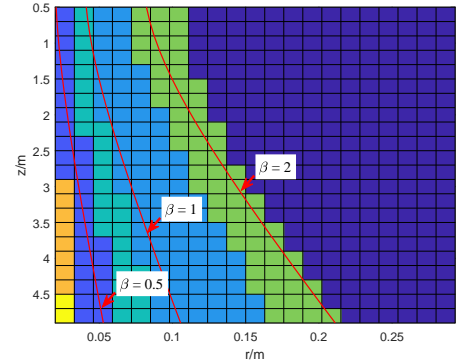}\label{dmin_map.d}
    }
	\caption{Multi-dimensional constellation maps with $I=2$, $f_1 = 60$ GHz. }\label{dmin_map}
\end{figure}

Then, we focus on the system with total power constraint, and gives some examples to study the properties of multi-dimensional constellation maps. Figure \ref{dmin_map} shows the multi-dimensional constellation maps with different system parameters. Different colors represent different categories, and it is observed that the OAM beam regions are classified into multiple categories. Each red curve in Figure \ref{dmin_map} represents all the positions ($\beta r_{a,max}^{l_m}(z), z$) with a fixed $\beta$, and it is seen that $\beta$ can determine the horizontal distance from the beam axis $(r=0)$. We set $\lambda_a = \lambda_1$ and $l_m = +1$. It is seen that the region in each category diverge along the same direction as $z$ increases, and in fact, this direction can be approximated as $r_{a,max}^{l_m}(z)$. According to the analysis above Theorem \ref{theorem.dmin_diff_nonuniform_linkgain}, the link gain ratio between any two sub-channels remains approximately unchanged for any positions ($\beta r_{a,max}^{l_m}(z), z$) with fixed $\beta$, and thus the link gain matrices are nearly proportional, which further leads to a fixed constellation at these positions. Hence, Figure \ref{dmin_map} verifies Theorem \ref{theorem.dmin_diff_nonuniform_linkgain}. 

In Figure \ref{dmin_map.a}, we consider the OAM mode set $\mathcal{L}=\{0, +1\}$, two carrier frequency $f_1 = 60$ GHz and $f_2 = 61$ GHz. When $l = +1$ and $i = 1$, $\beta^{+1}_{1,max} = \sqrt{\frac{\lambda_1 |+1|}{\lambda_a |l_m|} } = 1$, and therefore the curve $\beta = 1$ represents the maximum link gain positions of the sub-channel with OAM mode $+1$ in the first sub-system. When $l = +1$ and $i = 2$, the maximum link gain positions almost overlap the curve $\beta = 1$ due to the small difference between $f_1$ and $f_2$. We also present the curves $\beta = 0.4$ and $\beta = 2$ in Figure \ref{dmin_map.a}, which represent the positions of the left and right boundary regions, respectively. The results shows that $\beta $ can characterize the horizontal distance better than $r$. According to the analysis in Section \ref{subsection.Robustness_high_dim_cons}, it is easy to know that compared with the region $\beta < 1$, the channel matrix changes more slowly within $\beta > 1$. Hence, there are more categories in region $\beta < 1$ than those in region $\beta > 1$. Figures \ref{dmin_map.b}, \ref{dmin_map.c} and \ref{dmin_map.d} reveal similar features as those in Figure \ref{dmin_map.a}. In Figure \ref{dmin_map.b}, $\beta=1.414$ represents the maximum link gain positions $\beta^{+2}_{1,max}$. Compared with Figure \ref{dmin_map.a}, it is seen that the map in Figure \ref{dmin_map.b} has more categories at the positions with small $r$. The reason is that the OAM beams with modes $\pm2$ have a larger beam divergence than the OAM beams with modes $\pm1$, so the channel condition changes rapidly in this region. Figure \ref{dmin_map.c} shows the constellation map with two carrier frequency $f_1 = 60$ GHz and $f_2 = 65$ GHz, wherein $\beta=1.04$ represents the maximum link gain positions $\beta^{+1}_{2,max}$. Compared with Figure \ref{dmin_map.a}, the map in Figure \ref{dmin_map.c} has slightly more categories. The reason is that when $\triangle f=5$ GHz, the beam divergence of different carrier frequencies is slightly different, even if the OAM mode is the same. Figure \ref{dmin_map.d} shows the constellation map with $M=32$. It is seen that the region in each category diverges along the direction of a certain $\beta $, and the number of different categories in region $\beta > 1$ is larger than that in region $\beta < 1$. This confirms that the classification of a multi-dimensional constellation map mainly depends on positions rather than the modulation order.

\begin{figure}[htbp]
	\centering
	\subfloat[$\mathcal{L}=\{0, +1\}$, $\triangle f=1$ GHz]{
		\includegraphics[width=2.7 in] {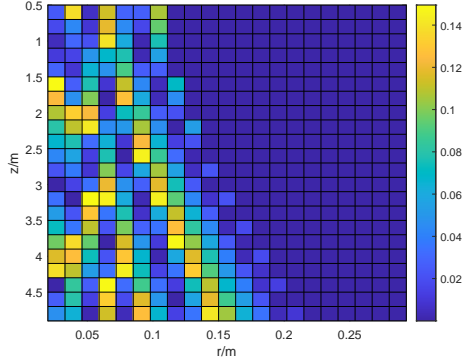}\label{nMED_Diff.a}
	}

	\subfloat[$\mathcal{L}=\{0,+2\}$, $\triangle f=1$ GHz]{
		\includegraphics[width=2.7 in] {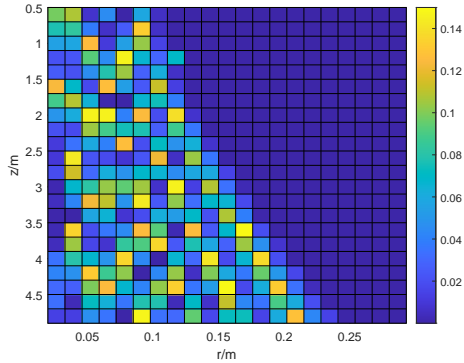}\label{nMED_Diff.b}
	}
	\caption{The normalized MED differences between using optimal constellation and mapped constellation. }\label{nMED_Diff}
\end{figure}

Figures \ref{nMED_Diff.a} and \ref{nMED_Diff.b} show the normalized MED differences of the multi-dimensional constellation maps in Figure \ref{dmin_map.a} and \ref{dmin_map.b}, respectively. It is observed that the normalized MED differences are irregular in the whole OAM beam region. The reason is that the normalized MED differences are not uniformly continuous with respect to $\|\Delta \mathbf{H} \|_F$, which means that the normalized MED differences cannot be uniformly bounded even if the $\Delta \mathbf{H}$ are the same at different positions. Furthermore, the solution of the approximated optimization problem using iterative algorithm may be not global optimal. Hence, the normalized MED difference may fluctuate within the OAM beam region. This fluctuation can also be observed in Figures \ref{diff_op_ep_01} and \ref{diff_op_ep_02}.

In summary, for short-range LOS links, multi-dimensional constellation map can be constructed as an offline look-up table to assign the constellation based on the position. Although there are multiple categories within the region with small $r$, the system can adopt a fixed multi-dimensional constellation at some positions with better channel condition. However, it is worth mentioning that as the number of OAM modes and wavelengths increases, the multi-dimensional constellation map will become more complicated. Hence, multi-dimensional constellation map can be adopted only when the number of sub-channels is small.

\section{conclusion} \label{section.conclusion}
In this paper, we consider a mmWave WDM system with positioning information in short-range LOS environment. We also consider the OAM multiplexing in each wavelength part. Based on the quasi-static feature of the communication links, we have proposed a map-assisted approach to offline store the system parameters according to the transceiver's position and configure the system by employing these parameters when obtaining the position information. To reduce the storage and search complexity, we attempt to classify all considered positions into several categories and employ one fixed parameter in each category. We have applied the map-assisted method to the multi-dimensional constellation design problem. We have analyzed the features of OAM beams with different modes, and found that the link gain ratio between any two sub-channels remains unchanged at certain positions. Then, we have investigated the effect of channel condition on the multi-dimensional constellation design, and figured out that the normalized MED difference of fixed constellation is sufficiently small if link gain matrices are sufficiently close to be proportional. Moreover, we have also analyzed the effect of fixed-power allocation on the multi-dimensional constellation design, and found that the system adopting fixed-power allocation have a negligible performance loss if the difference between fixed-power vector and optimal power vector is small. Numerical results have verified the theoretical analysis.


\section{Acknowledegments}
The authors would like to thank Dr. Qian Gao (Futurewei Technology) for his constructive comments and suggestions that improve the quality of this paper.

\begin{appendices} 

		\section{ proof of Theorem \ref{theorem.dmin_diff_nonuniform_linkgain}} \label{appendix.dmin_diff_nonuniform}
	According to the definition of MED, we have
	\begin{align}  \label{equ.deltaH_upperbound}
		& 1 - \frac{d_{min}(\mathbf{H}^{(2)}, \mathcal{C}^{*,(1)} ) }{d_{min}(\mathbf{H}^{(2)}, \mathcal{C}^{*,(2)} )}  \notag \\
		= & 1 -\frac{\| \mathbf{H}^{(2)}   (\mathbf{x}_2^{(1)}-\tilde{\mathbf{x}}_2^{(1)})\|_2 }{\|\mathbf{H}^{(2)} (\mathbf{x}_2^{(2)}-\tilde{\mathbf{x}}_2^{(2)})\|_2}  \notag \\
		= & 1 -\frac{\|(\alpha \mathbf{H}^{(1)} + \Delta \mathbf{H}) (\mathbf{x}_2^{(1)}-\tilde{\mathbf{x}}_2^{(1)})\|_2 }{\|\mathbf{H}^{(2)} (\mathbf{x}_2^{(2)}-\tilde{\mathbf{x}}_2^{(2)})\|_2}  \notag \\
		\leq & 1 -\frac{\| \alpha \mathbf{H}^{(1)}  (\mathbf{x}_2^{(1)}-\tilde{\mathbf{x}}_2^{(1)})\|_2 - \| \Delta \mathbf{H}  (\mathbf{x}_2^{(1)}-\tilde{\mathbf{x}}_2^{(1)})\|_2}{\|\mathbf{H}^{(2)}  (\mathbf{x}_2^{(2)}-\tilde{\mathbf{x}}_2^{(2)})\|_2}  \notag \\
		\leq & 1 -\frac{\| \alpha \mathbf{H}^{(1)}  (\mathbf{x}_1^{(1)}-\tilde{\mathbf{x}}_1^{(1)})\|_2 - \| \Delta \mathbf{H} (\mathbf{x}_2^{(1)}-\tilde{\mathbf{x}}_2^{(1)})\|_2}{\|\mathbf{H}^{(2)}  (\mathbf{x}_2^{(2)}-\tilde{\mathbf{x}}_2^{(2)})\|_2}  \notag \\
		\leq  & 1 -\frac{\| \alpha \mathbf{H}^{(1)}  (\mathbf{x}_1^{(2)}-\tilde{\mathbf{x}}_1^{(2)})\|_2 - \| \Delta \mathbf{H} (\mathbf{x}_2^{(1)}-\tilde{\mathbf{x}}_2^{(1)})\|_2}{\|\mathbf{H}^{(2)}  (\mathbf{x}_2^{(2)}-\tilde{\mathbf{x}}_2^{(2)})\|_2}  \notag \\
		\leq & 1 - \frac{\| \mathbf{H}^{(2)}   (\mathbf{x}_1^{(2)}-\tilde{\mathbf{x}}_1^{(2)})\|_2}{\|\mathbf{H}^{(2)}  (\mathbf{x}_2^{(2)}-\tilde{\mathbf{x}}_2^{(2)})\|_2} \notag \\
		& \quad + \frac{\| \Delta \mathbf{H}  (\mathbf{x}_1^{(2)}-\tilde{\mathbf{x}}_1^{(2)})\|_2 + \| \Delta \mathbf{H} (\mathbf{x}_2^{(1)}-\tilde{\mathbf{x}}_2^{(1)})\|_2}{\|\mathbf{H}^{(2)}  (\mathbf{x}_2^{(2)}-\tilde{\mathbf{x}}_2^{(2)})\|_2} \notag \\
		\leq & 1 - \frac{\| \mathbf{H}^{(2)}   (\mathbf{x}_2^{(2)}-\tilde{\mathbf{x}}_2^{(2)})\|_2}{\|\mathbf{H}^{(2)}  (\mathbf{x}_2^{(2)}-\tilde{\mathbf{x}}_2^{(2)})\|_2} \notag \\
		& \quad + \frac{\| \Delta \mathbf{H}  (\mathbf{x}_1^{(2)}-\tilde{\mathbf{x}}_1^{(2)})\|_2 + \| \Delta \mathbf{H} (\mathbf{x}_2^{(1)}-\tilde{\mathbf{x}}_2^{(1)})\|_2}{\|\mathbf{H}^{(2)}  (\mathbf{x}_2^{(2)}-\tilde{\mathbf{x}}_2^{(2)})\|_2} \notag \\
		\leq & \frac{\| \Delta \mathbf{H}  (\mathbf{x}_1^{(2)}-\tilde{\mathbf{x}}_1^{(2)})\|_2 + \| \Delta \mathbf{H} (\mathbf{x}_2^{(1)}-\tilde{\mathbf{x}}_2^{(1)})\|_2}{\|\mathbf{H}^{(2)}  (\mathbf{x}_2^{(2)}-\tilde{\mathbf{x}}_2^{(2)})\|_2}  \notag \\
		\leq & \frac{ \|\Delta \mathbf{H} \|_F(\| \mathbf{x}_1^{(2)}-\tilde{\mathbf{x}}_1^{(2)}\|_2 + \|\mathbf{x}_2^{(1)}-\tilde{\mathbf{x}}_2^{(1)}\|_2)}{\|\mathbf{H}^{(2)}  (\mathbf{x}_2^{(2)}-\tilde{\mathbf{x}}_2^{(2)})\|_2}.
	\end{align}
	Since $\mathbf{x}_2^{(2)}$ and $\tilde{\mathbf{x}}_2^{(2)}$ are two different constellations in $\mathcal{C}^{*,(2)}$, the term $ \|\mathbf{H}^{(2)}  (\mathbf{x}_2^{(2)}-\tilde{\mathbf{x}}_2)\|_2$ is non-zero. Besides, the terms $\|\mathbf{x}_1^{(2)}-\tilde{\mathbf{x}}_1^{(2)}\|_2 $ and $ \| \mathbf{x}_2^{(1)}-\tilde{\mathbf{x}}_2^{(1)}\|_2$ are generally finite. Hence, term $(\|\mathbf{x}_1^{(2)}-\tilde{\mathbf{x}}_1^{(2)}\|_2 + \| \mathbf{x}_2^{(1)}-\tilde{\mathbf{x}}_2^{(1)}\|_2) / \|\mathbf{H}^{(2)} (\mathbf{x}_2^{(2)}-\tilde{\mathbf{x}}_2^{(2)})\|_2$ is finite.

	\section{ proof of Theorem \ref{theorem.dmin_diff_power_allocation}} \label{appendix.dmin_diff}
	Letting $\mathbf{A}^{(o)}=\sqrt{\text{diag}(\mathbf{p}^{(o)})}$ and $\mathbf{A}^{(f)}=\sqrt{\text{diag}(\mathbf{p}^{(f)})}$, we first consider the following normalized MED difference, 
	\begin{align} \label{equ.pfSfpoSf}
	&\left| 1 - \frac{d_{min}(\mathbf{H}, \mathbf{p}^{(f)}, \mathcal{S}^{(f)} ) }{d_{min}(\mathbf{H}, \mathbf{p}^{(o)}, \mathcal{S}^{(f)} )}  \right| 
	\leq  \left| 1 - \frac{d_{min}^2(\mathbf{H}, \mathbf{p}^{(f)}, \mathcal{S}^{(f)} ) }{d_{min}^2(\mathbf{H}, \mathbf{p}^{(o)}, \mathcal{S}^{(f)} )}  \right|  \notag \\
	= & \left| 1 - \frac{\|\mathbf{H}\mathbf{A}^{(f)}(\mathbf{s}_f^{(f)}-\tilde{\mathbf{s}}_f^{(f)})\|_2^2 }{\|\mathbf{H}\mathbf{A}^{(o)}(\mathbf{s}_o^{(f)}-\tilde{\mathbf{s}}_o^{(f)})\|_2^2}  \right| \notag \\
	\leq & \max \Bigg\{\left| 1 - \frac{\|\mathbf{H}\mathbf{A}^{(f)}(\mathbf{s}_f^{(f)}-\tilde{\mathbf{s}}_f^{(f)})\|_2^2 }{\|\mathbf{H}\mathbf{A}^{(o)}(\mathbf{s}_f^{(f)}-\tilde{\mathbf{s}}_f^{(f)})\|_2^2}  \right|, \notag \\ & \quad \quad \quad \left| 1 - \frac{\|\mathbf{H}\mathbf{A}^{(f)}(\mathbf{s}_o^{(f)}-\tilde{\mathbf{s}}_o^{(f)})\|_2^2 }{\|\mathbf{H}\mathbf{A}^{(o)}(\mathbf{s}_o^{(f)}-\tilde{\mathbf{s}}_o^{(f)})\|_2^2}  \right| \Bigg\},
	\end{align}
	where
	\begin{align}
	&\left| 1 - \frac{\|\mathbf{H}\mathbf{A}^{(f)}(\mathbf{s}_f^{(f)}-\tilde{\mathbf{s}}_f^{(f)})\|_2^2 }{\|\mathbf{H}\mathbf{A}^{(o)}(\mathbf{s}_f^{(f)}-\tilde{\mathbf{s}}_f^{(f)})\|_2^2}  \right| \notag \\
	=  &\frac{\left| \sum_{i =1 }^N \frac{|h_i|^2}{N_i}(p_i^{(o)} - p_i^{(f)})|\mathbf{s}_f^{(f)}(i)-\tilde{\mathbf{s}}_f^{(f)}(i)|^2 \right| }{\|\mathbf{H}\mathbf{A}^{(o)}(\mathbf{s}_f^{(f)}-\tilde{\mathbf{s}}_f^{(f)})\|_2^2} \notag \\
	\leq & \frac{ \sum_{i =1 }^N \frac{|h_i|^2}{N_i} |p_i^{(o)} - p_i^{(f)}|\cdot |\mathbf{s}_f^{(f)}(i)-\tilde{\mathbf{s}}_f^{(f)}(i)|^2  }{\|\mathbf{H}\mathbf{A}^{(o)}(\mathbf{s}_f^{(f)}-\tilde{\mathbf{s}}_f^{(f)})\|_2^2} \notag \\
	\leq & \frac{ \left( \sum_{i =1 }^N \frac{|h_i|^4}{N_i^2}  |\mathbf{s}_f^{(f)}(i)-\tilde{\mathbf{s}}_f^{(f)}(i)|^4\right) ^\frac{1}{2} \cdot \left( \sum_{i =1 }^N |p_i^{(o)} - p_i^{(f)}|^2 \right) ^\frac{1}{2} }{\|\mathbf{H}\mathbf{A}^{(o)}(\mathbf{s}_f^{(f)}-\tilde{\mathbf{s}}_f^{(f)})\|_2^2} \notag \\
	\leq & \frac{  (\sum_{i =1 }^N \frac{|h_i|^2}{N_i}  |\mathbf{s}_f^{(f)}(i)-\tilde{\mathbf{s}}_f^{(f)}(i)|^2 ) \cdot \left\| \mathbf{p}^{(o)} - \mathbf{p}^{(f)} \right\|_2  }{\|\mathbf{H}\mathbf{A}^{(o)}(\mathbf{s}_f^{(f)}-\tilde{\mathbf{s}}_f^{(f)})\|_2^2} \notag \\
	= & \frac{  \|\mathbf{H}(\mathbf{s}_f^{(f)}-\tilde{\mathbf{s}}_f^{(f)})\|_2^2 \cdot \left\| \mathbf{p}^{(o)} - \mathbf{p}^{(f)} \right\|_2  }{\|\mathbf{H}\mathbf{A}^{(o)}(\mathbf{s}_f^{(f)}-\tilde{\mathbf{s}}_f^{(f)})\|_2^2}.
	\end{align}
	Similarly, we can obtain the upper bound of $\left| 1 - \|\mathbf{H}\mathbf{A}^{(f)}(\mathbf{s}_o^{(f)}-\tilde{\mathbf{s}}_o^{(f)})\|_2^2 /\|\mathbf{H}\mathbf{A}^{(o)}(\mathbf{s}_o^{(f)}-\tilde{\mathbf{s}}_o^{(f)})\|_2^2  \right| $. 
	Hence, the upper bound of $\left| 1 - d_{min}(\mathbf{H}, \mathbf{p}^{(f)}, \mathcal{S}^{(f)} ) /d_{min}(\mathbf{H}, \mathbf{p}^{(o)}, \mathcal{S}^{(f)} )  \right|$ can be written as
	\begin{align} \label{equ.pfSfpoSo}
	&\left| 1 - \frac{d_{min}(\mathbf{H}, \mathbf{p}^{(f)}, \mathcal{S}^{(f)} ) }{d_{min}(\mathbf{H}, \mathbf{p}^{(o)}, \mathcal{S}^{(f)} )}  \right| \leq \left\| \mathbf{p}^{(o)} - \mathbf{p}^{(f)} \right\|_2 \cdot \notag \\
	&\quad \max \left\{ \frac{  \|\mathbf{H}(\mathbf{s}_f^{(f)}-\tilde{\mathbf{s}}_f^{(f)})\|_2^2}{\|\mathbf{H}\mathbf{A}^{(o)}(\mathbf{s}_f^{(f)}-\tilde{\mathbf{s}}_f^{(f)})\|_2^2}, \frac{  \|\mathbf{H}(\mathbf{s}_o^{(f)}-\tilde{\mathbf{s}}_o^{(f)})\|_2^2}{\|\mathbf{H}\mathbf{A}^{(o)}(\mathbf{s}_o^{(f)}-\tilde{\mathbf{s}}_o^{(f)})\|_2^2}\right\}.
	\end{align}
	Using similar steps in Equations \eqref{equ.pfSfpoSf}-\eqref{equ.pfSfpoSo}, we obtain
	\begin{align} \label{equ.pfSopoSo}
		&\left| 1 - \frac{d_{min}(\mathbf{H}, \mathbf{p}^{(f)}, \mathcal{S}^{(o)} ) }{d_{min}(\mathbf{H}, \mathbf{p}^{(o)}, \mathcal{S}^{(o)} )}  \right| \leq \left\| \mathbf{p}^{(o)} - \mathbf{p}^{(f)} \right\|_2 \cdot \notag \\
		&\quad \max \left\{ \frac{  \|\mathbf{H}(\mathbf{s}_o^{(o)}-\tilde{\mathbf{s}}_o^{(o)})\|_2^2}{\|\mathbf{H}\mathbf{A}^{(o)}(\mathbf{s}_o^{(o)}-\tilde{\mathbf{s}}_o^{(o)})\|_2^2}, \frac{  \|\mathbf{H}(\mathbf{s}_f^{(o)}-\tilde{\mathbf{s}}_f^{(o)})\|_2^2}{\|\mathbf{H}\mathbf{A}^{(o)}(\mathbf{s}_f^{(o)}-\tilde{\mathbf{s}}_f^{(o)})\|_2^2}\right\}.
	\end{align}
	
	Note that the normalized MED difference $\left| 1 - d_{min}(\mathbf{H}, \mathbf{p}^{(f)}, \mathcal{S}^{(f)} ) / d_{min}(\mathbf{H}, \mathbf{p}^{(o)}, \mathcal{S}^{(o)} ) \right|$ satisfies
	\begin{align}
	&\left| 1 - \frac{d_{min}(\mathbf{H}, \mathbf{p}^{(f)}, \mathcal{S}^{(f)} ) }{d_{min}(\mathbf{H}, \mathbf{p}^{(o)}, \mathcal{S}^{(o)} )}  \right| \leq \notag \\
	& \max \left\{ \left| 1 - \frac{d_{min}(\mathbf{H}, \mathbf{p}^{(f)}, \mathcal{S}^{(f)} ) }{d_{min}(\mathbf{H}, \mathbf{p}^{(o)}, \mathcal{S}^{(f)} )}  \right|, \left| 1 - \frac{d_{min}(\mathbf{H}, \mathbf{p}^{(f)}, \mathcal{S}^{(o)} ) }{d_{min}(\mathbf{H}, \mathbf{p}^{(o)}, \mathcal{S}^{(o)} )}  \right| \right\}.
	\end{align}
	Combining \eqref{equ.pfSfpoSo} and \eqref{equ.pfSopoSo}, it yields
	\begin{align}
	&\left| 1 - \frac{d_{min}(\mathbf{H}, \mathbf{p}^{(f)}, \mathcal{S}^{(f)} ) }{d_{min}(\mathbf{H}, \mathbf{p}^{(o)}, \mathcal{S}^{(o)} )}  \right| \notag \\
	&\leq \left\| \mathbf{p}^{(o)} - \mathbf{p}^{(f)} \right\|_2 \cdot \max_{\mathbf{s}_d \in \mathcal{S}_d} \left\{ \frac{  \|\mathbf{H}\mathbf{s}_d\|_2^2}{\|\mathbf{H}\mathbf{A}^{(o)}\mathbf{s}_d\|_2^2}\right\},
	\end{align}
	where $\mathcal{S}_d$ is the set of $ \left\{ \mathbf{s}_o^{(o)}-\tilde{\mathbf{s}}_o^{(o)}, \mathbf{s}_o^{(f)}-\tilde{\mathbf{s}}_o^{(f)}, \mathbf{s}_f^{(o)}-\tilde{\mathbf{s}}_f^{(o)},  \mathbf{s}_f^{(f)}-\tilde{\mathbf{s}}_f^{(f)} \right\}$. Generally, the constellation points of $\mathcal{S}^{(f)}$ do not only distribute on the sub-channels with zero power, when $\mathbf{p}^{(f)}$ is set reasonably. Hence, the term $\|\mathbf{H}\mathbf{A}^{(o)}\mathbf{s}_d\|_2^2$ is non-zero and the term $\max_{\mathbf{s}_d \in \mathcal{S}_d} \left\{    \|\mathbf{H}\mathbf{s}_d\|_2^2 / \|\mathbf{H}\mathbf{A}^{(o)}\mathbf{s}_d\|_2^2\right\}$ is finite.

\end{appendices} 

\begin{footnotesize}
	\bibliographystyle{IEEEtran}
	\bibliography{bib/Map_assisted_mmWave_WDM}
\end{footnotesize}

\end{document}